# A solution for applying IEC 61499 function blocks in the development of substation automation systems

Valentin Vlad, Cezar D. Popa, Corneliu O. Turcu, Corneliu Buzduga

*Abstract*—This paper presents a solution for applying IEC 61499 function blocks along with IEC 61850 specifications in modeling and implementing control applications for substations automation. The IEC 61499 artifacts are used for structuring the control logic, while the IEC 61850 concepts for communication and information exchange between the automation devices. The proposed control architecture was implemented and validated in a simple fault protection scenario with simulated power equipment.

## I. Introduction

RECENT advances in substation automation systems include the adoption of a new standard for communication between intelligent electronic devices (IEDs), published by the International Electrotechnical Commission (IEC) under the name IEC 61850 [1], [2], [11], [12], [13]. The standard specifies how the data should be organized in IEDs, defines a list of services for reading and modifying these data, and proposes a mapping of the data and services to a specific communication protocol [1], [2]. However, the standard does not deal with the implementation of the control logic encapsulated in IEDs, nor how different control components in the same IED should communicate with each other, but rather with the external behavior of the automation devices. Different implementations for the communication interface defined by the standard exist, commercial or open source, in different programming languages, like C and Java. There are also proprietary implementations, included in commercial protection and control devices for automations in substations [3], [4].

A convenient solution for developing modular control applications is the one proposed by the IEC 61499 standard, focused on applying function blocks in the design of distributed industrial-process measurement and control systems. IEC 61499 includes advanced software technologies, such as the encapsulation of functionality, component-based design, event-driven execution and distribution, and was intended as support technology for the development of holonic manufacturing control systems [5].

In this paper we propose a control architecture for substation automation systems which combines the IEC 61499 models (for the implementation of the control logic) with the IEC 61850 specifications, for communication interface. The reasons for choosing the IEC 61499 standard for modeling the control logic are mainly related to its comprehensive methodology of building control applications from modular components (function blocks), in a graphical way. The scientific literature contains also published work on harmonizing the IEC 61850 and IEC 61499 standards, the most representative papers including [6], [7] and [8].

The rest of the document is organized as follows. The next chapter gives an overview of the IEC 61850 and IEC 61499 standards. The third chapter introduces the proposed control architecture, and the fourth one presents its implementation and validation in a simple fault protection scenario. The last chapter is dedicated to conclusions.

## II. IEC 61850 AND IEC 61499

### A. IEC 61850

IEC 61850 is a relatively new standard (2004) designed to improve the communication between the automation devices within substations. In contrast with the legacy protocols, focused on how bytes are transmitted on the wire, the standard deals also with the internal organization of data in devices, facilitating a better interoperability between them and reduced configuration costs [1].

The standard adopts a model-driven approach by standardizing device, object and service models. A physical automation device is typically defined by its network address and contains one or more logical devices. Each logical device contains one or more logical nodes (LN), which are named grouping of data and associated services, logically related to some power system function (Fig. 1). Different types of logical nodes are defined, e.g. for automatic control, for metering and

This work was supported in part by the project entitled „SOCERT. Knowledge society, dynamism through research", contract number POSDRU/159/1.5/S/132406.

V. Vlad, C.D. Popa, C.O. Turcu and C.Buzduga are with the Electrical Engineering Department and Computer Science, Stefan cel Mare University of Suceava, 13, Universitatii, Romania (corresponding author to provide phone: 0230-524801; e-mail: vladv@eed.usv.ro).

C. D. Popa, C.O. Turcu and C.Buzduga are with the Electrical Engineering Department and Computer Science, Stefan cel Mare University of Suceava, 13, Universitatii, Romania (e-mail: {cezarp | cturcu | cbuzduga}@eed.usv.ro).

measurement, for supervisory control, protection, switchgears, etc. The logical node instances have standardized names, consisting of a part of four letters (e.g. XCBR, for circuit breaker control) and a suffix, representing the instance ID (e.g. XCBR1, XCBR2). An optional, application specific prefix can also be used as part of the LN instance name. Each logical node contains one or more elements of Data, with names and types defined by the standard. Each element of Data within the logical node conforms to the specification of a common data class (CDC), describing the type and structure of the data. The data attributes of the common data classes contains in fact the data of the logical node. For example, an XCBR-type logical node will include an element of Data named Pos, containing an attribute stVal, whose value represents the position of the circuit breaker.

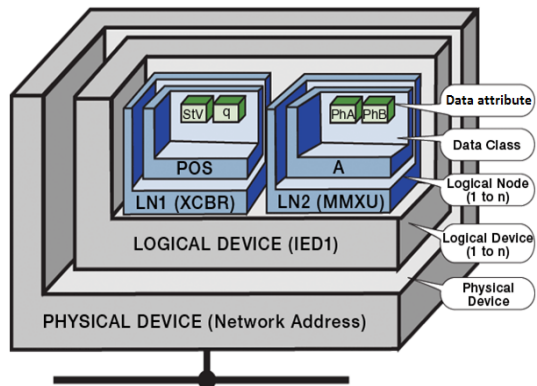

Fig. 1 Organization of data in IEDs according to IEC 61850 (from [9])

The communication services are defined in part 7-2 of the standard by the Abstract Communication Service Interface (ACSI). Two types of communication are defined: one based on a client-server mechanism, for reading and setting data in devices, and a peer to peer communication type for fast and reliable system-wide event distribution, based on a publisher-subscriber mechanism. Sampled measured values are also transmitted through this publisher-subscriber mechanism.

The abstract model of the device and the communication services are mapped to a specific communication protocol stack based on MMS (Manufacturing Messaging Specification – ISO 9506), TCP/IP, and Ethernet.

The capabilities of an IED can be described using an XML-based substation configuration description language – SCL, which enables the information exchange between engineering tools from different manufacturers.

As to the substation architecture, the standard places the monitor and control equipment on three levels, namely process, bay and substation level, as illustrated in Fig. 2. The devices at the process level are designed to collect information such as voltage, current, and status information from the primary equipment and to transmit them in a digital form to the upper level. The bay level includes IEDs running applications for protection and control, while the substation level is dedicated to applications for monitoring and control of the whole substation.

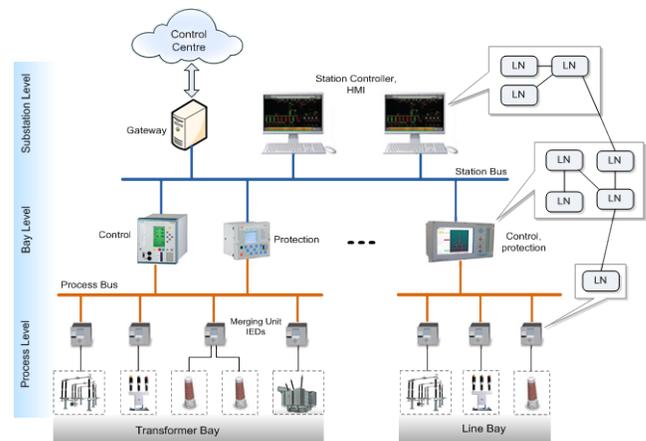

Fig. 2 Substation architecture according to the IEC 61850 standard

### B. IEC 61499

IEC 61499 defines an open architecture for distributed control and automation, presented in terms of implementable reference models, textual syntax and graphical representations [10]. The programming unit of the IEC 61499 is the function block (FB), from which complex control applications can be built. The standard defines three types of function blocks: basic, composite, and service interface function blocks. Basic function blocks perform elemental control functions, such as reading a sensor or setting the state of an actuator by executing various encapsulated algorithms, according to an execution control chart. The functionality of composite function blocks is determined by a network of interconnected function blocks inside. The service interface function blocks (SIFB) serve to abstract the specific functions of a hardware platform, allowing the application developer to focus on the control logic.

The control applications are built (according to IEC 61499) as networks of function blocks, which can be distributed across several devices. The IEC 61499 device model contains one or more resources, to which parts of a function block network can be mapped to be executed (Fig. 3).

According to [10], a resource is considered "a functional unit contained in a device, which has independent control of its operation. It may be created, parameterized, started up, deleted, etc., without affecting other resources within a device." This definition encouraged us to assimilate an IEC 61850 logical node to IEC 61499 resource, as explained in the next section.

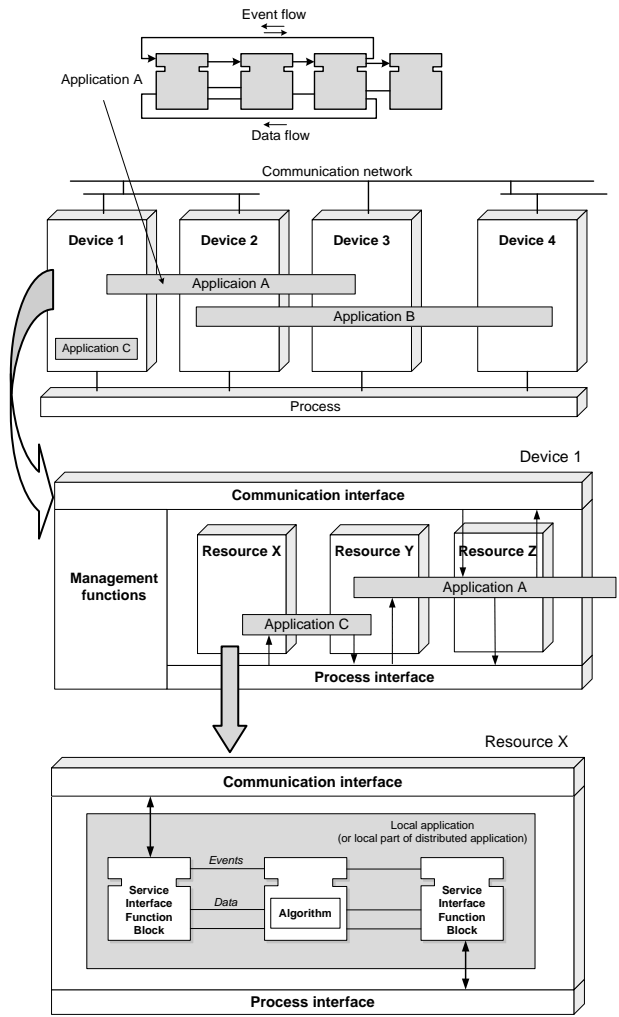

Fig. 3 Mapping of control applications to IEC 61499 devices and resources

### III. THE PROPOSED CONTROL ARCHITECTURE

As mentioned in introduction, our approach for implementing control applications for substation automation relies on IEC 61499 models for the control logic and on IEC 61850 specifications for communication and information exchange between IEDs. As illustrated in Fig. 4, we propose a logical node to be modeled and implemented as a network of function blocks mapped to an IEC 61499 resource, named according to the naming rules defined by the IEC 61850 standard for logical nodes. The function block network will include a special function block encapsulating the data of the logical node (as specified in the IEC 61850 standard) and function blocks for internal communications (with logical nodes in the same device) or for interfacing with the primary equipment. The special function block, in the following referred to as the main function block – MFB, will receive information from the primary equipment or from other logical nodes and will process it according to the encapsulated algorithms, updating its internal data and generating actions (e.g. sending event messages).

For communications between logical nodes in the same IED, the explicitly defined IEC 61499 communication function blocks based on the publish/subscribe mechanism can be used.

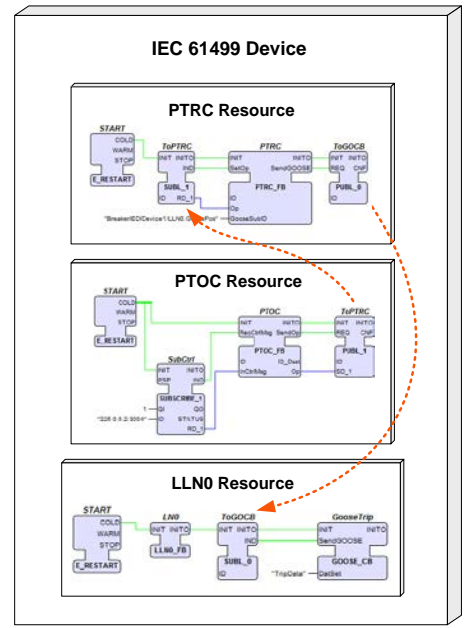

Fig. 4 Logical nodes modeled as networks of function blocks mapped to resources

As to the external behavior of the IED, an IEC 61850 server should be implemented as specified in part 7 2 of the standard. The server will exhibit outside the data indicated in the SCL configuration file, updated continuously from the MFBs of the logical nodes. The synchronization of data between the control logic and the server's buffers can be realized through special functions added to the IEC 61499 device containing the server and the logical nodes. These functions must allow the server to access the data contained in the MFBs, based on their IEC 61850-compliant references. This mechanism, along with the control architecture, is illustrated in Fig. 5, considering as example a Breaker IED.

The transmission of events as GOOSE (Generic Object Oriented Substation Events) messages or of the sampled values using multicast can be realized with specialized service interface function blocks, transmitting data at regular intervals or when requested by the MFBs.

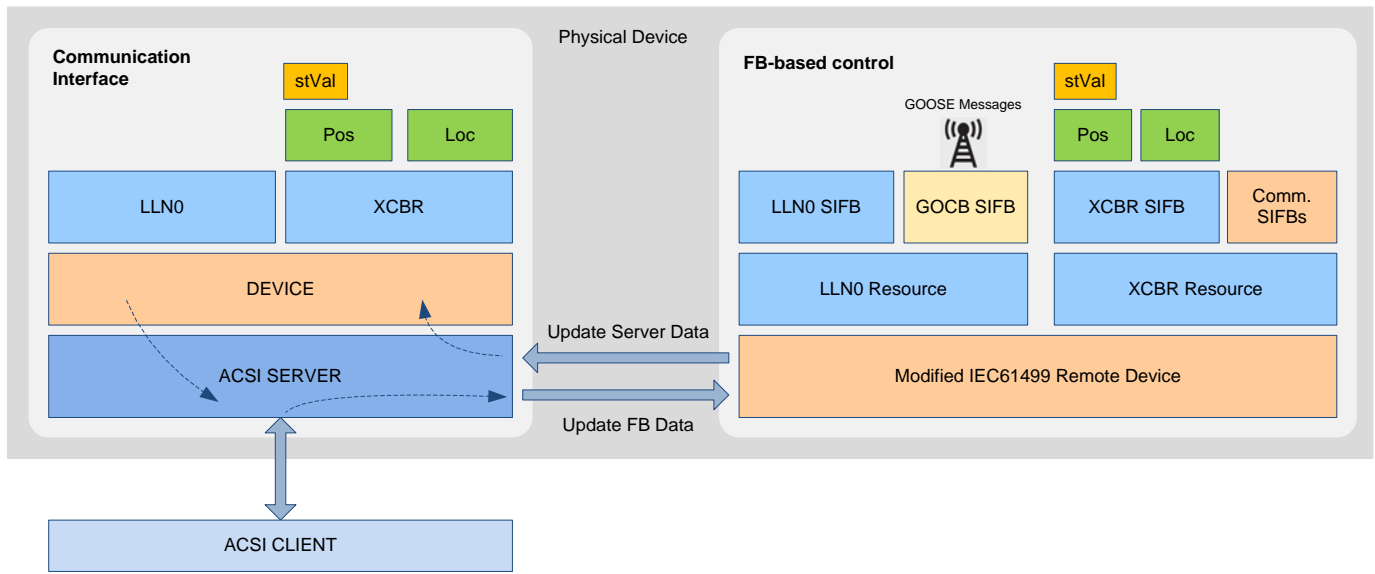

Fig.5. Proposed control and communication architecture (example for a Breaker IED)

## IV. IMPLEMENTATION ASPECTS

The proposed control architecture was implemented and tested in a simple fault protection scenario, with simulated power equipment.

The power system considered include a bus bar, a disconnector, a circuit breaker, a current transformer and a consumer, as illustrated in Fig. 6. The current through the feeder (the consumer load) can be set through a spinner.

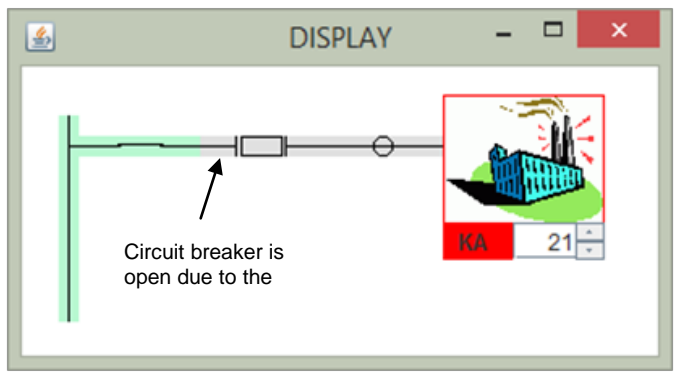

Fig.6. User interface simulating the power system

The protection scenario (illustrated in Fig. 7) involves the transmission of GOOSE messages between several IEC 61850 logical nodes for tripping the circuit breaker and for publishing its new position.

The values measured by the current transformer (CT) are transmitted to a PTOC (protection overcurrent) logical node. When an overcurrent is detected, PTOC communicates the anomalous condition to the PTRC (protection trip conditioning) LN, which issues a trip command (in form of a GOOSE message) to the XCBR LN. As a result the circuit breaker is open and the new status is transmitted (also through GOOSE messages) to the PTRC and RREC (auto-reclosing) logical nodes. After a short time, RREC issues a reclose command to the XCBR LN, which closes the circuit breaker and publishes its new status. If the abnormal condition persists, RREC will get blocked after three operations, the circuit breaker remaining open.

The control logic was implemented as an IEC 61499 system with five devices, as illustrated in Fig. 8. Four of the five devices models intelligent electronic devices encapsulating the control logic, while the fifth (DISPLAY) contains function blocks for simulating the power system equipment.

The function blocks modeling the logical nodes were encapsulated in IEC 61499 resources, with names compliant with the IEC 61850 specifications for logical node names. The MFBs of the logical nodes are implemented as service interface function blocks and communicate with the simulated power equipment and other logical nodes through FBs of Publish/Subscribe type. The Goose Control Block was modeled as a SIFB included in the resource of the LLN0 logical node.

Each element of the power system is simulated through a composite function block, including SIFBs for modeling the element's physical behavior and for communication with the control logic. Fig. 9 presents the network of FBs modeling the physical system and the content of the composite function block simulating the circuit breaker.

A special resource (called SERVER) was created in each device to run an IEC 61850 server for external communications. The data exhibited by the server is defined through an SCL file, which must be in accordance with the data implemented in the FB-based logical nodes. A third-party IEC 61850 client was used to examine the syncronization of data between the server side and the FB-based control logic.

For the implementation of the IEC 61850 server we used the Java open source library presented at [4].

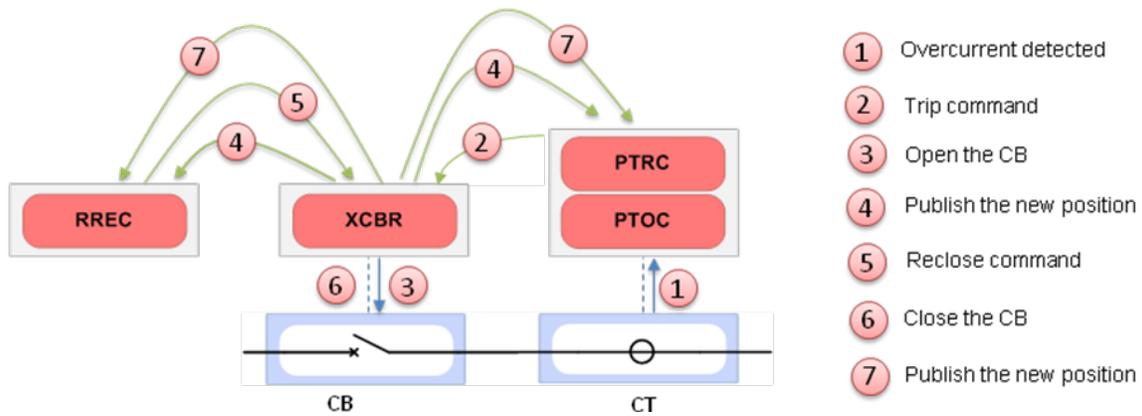

Fig.7. Fault protection scenario

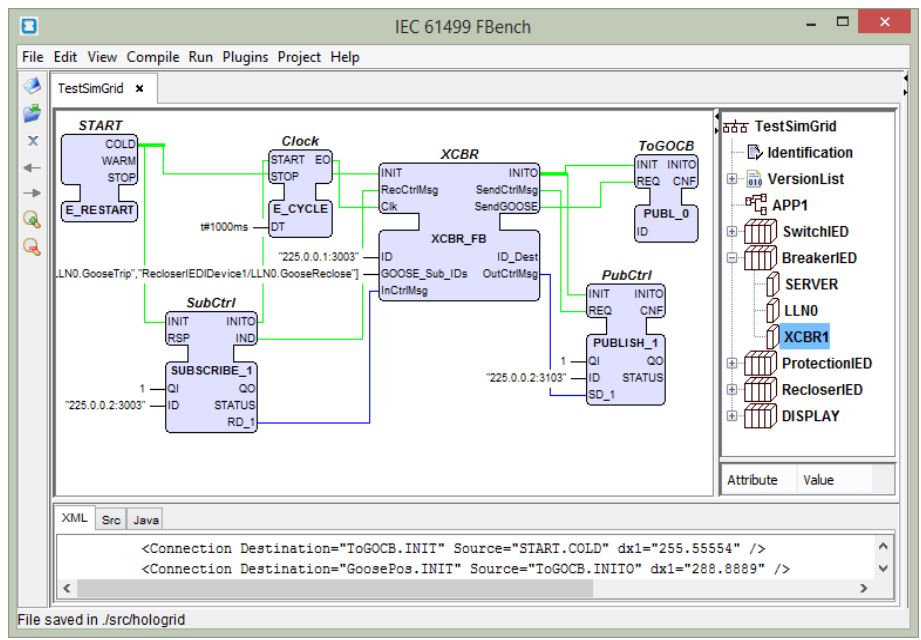

Fig.8. The IEC 61499 system implementing the control logic

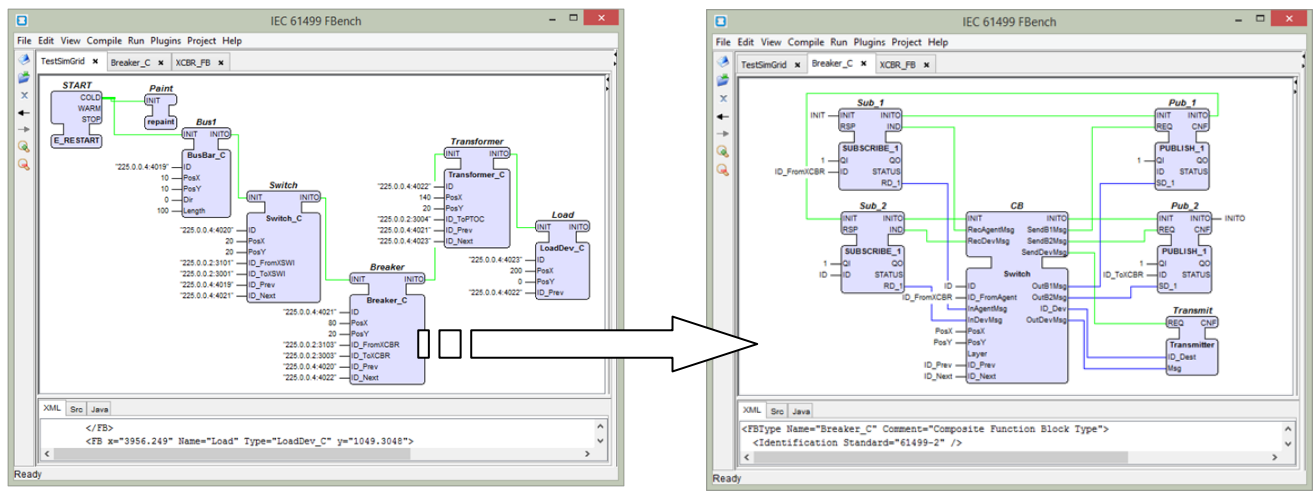

Fig.9. Solution for simulating the power system equipment with IEC 61499 FBs

## V. Conclusion

This paper presented a solution for applying IEC 61499 function blocks in modeling and implementing IEC 61850 – compliant control systems for substations automation. The use of function blocks allow for a comprehensible structuring of the control applications, modularity, and code reusability. Moreover, the IEC 61499 artifacts can easily accommodate the IEC 61850 models for the implementation of the IEDs control logic.

The proposed solution was implemented and tested in an overcurrent protection scenario with GOOSE messages, and the synchronization between the data on the server side and the control logic side was examined with a third-party IEC 61850 client.

### Acknowledgment

This paper has been financially supported within the project entitled *„SOCERT. Knowledge society, dynamism through research",* contract number POSDRU/159/1.5/S/132406. This project is co-financed by European Social Fund through Sectoral Operational Programme for Human Resources Development 2007-2013. **Investing in people!**"